\documentclass[12pt,preprint]{aastex}
\usepackage{epsfig}
\usepackage{natbib}

\usepackage{amsmath}
\usepackage{amssymb}
\usepackage{xspace}
\usepackage{color}

\def\msol{\mathrm{M}_\odot}

\def\enuc{\epsilon_{\rm nuc}}
\def\cv{c_{\rm V}}
\def\cp{c_{\rm P}}
\def\taunuc{\tau_{\rm nuc}}
\def\taud{\tau_{\rm d}}

\def\texpl{t_{\rm expl}}
\def\simgr{\,\hbox{\hbox{$ > $}\kern -0.8em \lower 1.0ex\hbox{$\sim$}}\,}
\def\simle{\,\hbox{\hbox{$ < $}\kern -0.8em \lower 1.0ex\hbox{$\sim$}}\,}
\def\beq{\begin{equation}}
\def\eeq{\end{equation}}
\def\sun{$_{\scriptscriptstyle \odot}$}

\shorttitle{Stability of SN Ia progenitors against radial oscillations}
\shortauthors{Baraffe et al.}

\begin{document}

\title{Stability of SN Ia progenitors against radial oscillations}

\author{I. Baraffe}
\affil{Ecole Normale Sup\'erieure, C.R.A.L, 46 all\'ee d'Italie,  Lyon, France}
\email{ibaraffe@ens-lyon.fr}

\author{A. Heger}
\affil{Theoretical Astrophysics Group, T-6, MS B227, Los Alamos National
Laboratory, Los Alamos, NM 87545} \email{alex@t6.lanl.gov}

\affil{Enrico Fermi Institute, The University
of Chicago, 5640 S.\ Ellis Ave, Chicago, IL 60637}

\and
\author{S.~E.\ Woosley}
\affil{Department of Astronomy and Astrophysics 
University of California, Santa Cruz, CA 95064, U.S.A.}
\email{woosley@ucolick.org}

\begin{abstract}
We analyze the possible existence of a pulsational instability excited
by the $\epsilon$-mechanism during the last few centuries of evolution
of a Chandrasekhar mass white dwarf prior to its explosion as a Type Ia
supernova. Our analysis is motivated by the temperature sensitivity of
the nuclear energy generation rate ($\sim T^{23}$) in a white dwarf
whose structural adiabatic index is near 4/3.  Based upon a linear
stability analysis, we find that the fundamental mode and higher order
radial modes are indeed unstable and that the fundamental mode has the
shortest growth time scale.  However, the growth time scale for such
instability never becomes shorter than the evolutionary timescale.
Therefore, even though the star \emph{is} pulsationally unstable, we
do not expect these radial modes to have time to grow and to affect
the structure and explosion properties of Type Ia supernovae.
\end{abstract}



\section{Introduction}

The explosion of a Chandrasekhar mass white dwarf as a Type Ia
supernova is one of the most dramatic events that can befall a star
and, at the same time, one of the most complex. The outcome depends on
details of the ignition process (Woosley, Wunsch, \& Kuhlen 2004) and
the uncertain physics of turbulent flame propagation (see Hillebrandt
\& Niemeyer 2000, Gamezo et al. 2003, and references therein).  Both
of these depend on the presupernova evolution of the star which sets
e.g., the central density at the time of runaway and the distribution
of ignition points. However, since pioneering studies in the 1970's by
Paczynski (1972), Couch \& Arnett (1973, 1974, 1975), and Iben (1978),
comparatively little attention has been given to the evolution of the
interior of the {\sl presupernova} star. It is known that the star
ignites carbon centuries before it finally explodes, commencing a long
ramp up that eventually concludes when convection can no longer carry
away the excess energy generated by the degenerate nuclear runaway.
It is during this phase that the URCA process may lead to a complex
convective structure in the core and perhaps even oscillations related
to the switching on and off of URCA neutrino losses (Iben 1978). In
this paper we consider a comparatively simple question: ``Is
presupernova carbon burning inherently unstable - even in the absence
of URCA losses''?

Carbon burning occurs at a rate that is extremely temperature
sensitive, at least T$^{23}$ even at the end of the stable burning
near $T \approx 7 \times 10^8$ K. The white dwarf is also supported
chiefly by degenerate electrons that are relativistic, hence $\Gamma
\approx 4/3$. In other contexts in stellar evolution, this combination
often tends towards instability (e.g., Ledoux 1941; Maeder 1985;
Baraffe, Heger, \& Woosley 2001).  Large excursions in the radius are
not strongly damped and the non-linearity of the reaction rate
promotes instability. This is known as the $\epsilon$-mechanism (Unno
et al.~1989).  Such an instability leads to oscillations around the
hydrostatic equilibrium configuration and periodic phases of expansion
and contraction on a dynamical timescale $\tau_{\rm dyn} \sim (G \bar
\rho)^{-1/2}$.  These types of oscillations are described as acoustic
modes.  They can be excited if an excitation mechanism operates in
regions where the amplitude of eigenfunctions is large.

We have thus undertaken a linear stability analysis of the structure
of Chandrasekhar mass white dwarf models prior to explosion, based on
a similar analysis performed for very massive stars (Baraffe et al.\
2001).  We pick up the white dwarf when it has just ignited carbon
(nuclear energy generation substantially in excess of plasma neutrino
losses) and is developing an extended convective core. The central
temperature at this point is $\sim3\times 10^8\,$K and it is about 300
years until the star's death.  The evolutionary models and input
physics used are described in \S 2 and the results of the linear
stability analysis are presented in \S 3.  A discussion follows in
\S4.

\section{Evolutionary models prior to explosion}

\subsection{Characteristic evolutionary timescales}

Following ignition, a convective region develops in the white dwarf
that eventually grows to encompass most of its mass. Heat is
transported outwards, but both radiative losses and neutrino losses
are negligible.  Instead, the energy mostly goes into heating the
convective portion of the star, and, to a lesser extent, expansion.
The heat capacity is given by
\begin{equation}
\begin{split}
  c_{\mathrm{P}} \ &= \ \left( \frac{\partial e_{\rm ions}}{\partial
      T} \right)_{\!\!P} +\left( \frac{\partial
      e_{\rm electrons}}{\partial T} \right)_{\!\!P} \\
  &= \ \left[9.1 \times 10^{14} \ + \ \frac{8.6 \times 10^{13} \ 
    T_8}{\rho_9^{1/3}}\right] \ \ 
\left(\frac{\mathrm{erg}}{10^8\,\mathrm{g\,K}}\right)
\label{heatcap}
\end{split}
\end{equation}
(Woosley et al. 2004), where $e$ is the specific internal energy, $T_8
=$ T/$10^8\,$K and $\rho_9 =\rho$/$10^9\,$g$\,$cm$^{-3}$.  For the
relevant temperature and density range, $T_8$ = 2 to 7, $\rho_9 = 1$
to 3, the ionic term dominates with a minor contribution from the
electrons.  To a factor-of-two accuracy, $\cp \approx 10^{15}$ erg
g$^{-1}$ (10$^8\,$K)$^{-1}$.  The total thermal energy is
\begin{equation}
H \ = \ \int_{0}^R \left[ \int_0^{T(r)} \ \cp(\rho(r),T) \, \rho(r) \, dT
\right] \ 4 \pi r^2 \ dr
\end{equation}




The correct definition for pure ``thermal energy'' should be expressed
in terms of $\cv$ rather than $\cp$, since $\cp$ also contains an
expansion term $PdV$.  However, because of strong degeneracy in the
interior of the white dwarf, $\cv \approx \cp$ and Eq.~(2) provides a
good estimate of the thermal heat content.  Moreover, for our purpose
below, to estimate the gravothermal heat content, this is the
appropriate quantity to use.  If the mass-averaged temperature is
about half the central value and the convection zone, about one solar
mass,
\begin{equation}
H \ \approx \ 7 \times 10^{48} \ \frac{T_{8,\rm c}}{7} \frac{M_{\rm
conv}} {M_{\scriptscriptstyle \odot}} \ \ {\rm erg}\,,
\end{equation}
where the index ``c'' stands for central quantities of the star.
Actually, from computer models discussed in the next section we know
that the convective core grows from 0.2 M\sun \ to 1.15 M\sun \ as the
central temperature rises from $3 \times 10^8$ K to $7 \times 10^8$ K
(see Table \ref{tab1}).  A better estimate of the heat in the
convective region is
\begin{equation}
H \ \approx \ 7 \times 10^{48} \ \left(\frac{T_{8,\rm c}}{7}\right)^3 \ \ 
{\rm erg}.
\end{equation}
This actually agrees to much better than factor-of-two accuracy with
the detailed computer results described in the next section.

The total nuclear power generated by the star is (Woosley et al. 2004)
\begin{equation}
L \ \approx \ 7.0 \times 10^{44} \ {\rm erg \ s^{-1}} \
(\frac{\rho_{9,\rm c}}{2})^{4.3} \, (\frac{T_{8,\rm c}}{7})^{\nu} \,
(\frac{X_{12}}{0.5})^2,
\label{loft}
\end{equation}
where $\nu \approx 23$ for the relevant range of temperature and
$X_{12}$ is the carbon mass fraction. This expression is valid in the
temperature range $4 \lesssim T_8 \lesssim 7$ and becomes increasingly
inaccurate at lower temperature because of temperature dependence of
$\nu$ and especially the complex dependence of the screening
correction on temperature and density.

The nuclear timescale, which characterizes the increase of nuclear
energy generation rate in the center, taking into account the reaction of
the rest of the star,
is
\begin{equation}
\begin{split}
\tau_{\rm L} \ &= \ \left(\frac{\mathrm{d}\ln L}{\mathrm{d}t}\right)^{\!\!-1} =
\ \frac{3H}{\nu \, L}\\
& \approx {\frac{3\, 10^4}{\nu}} \left(\frac{2}{\rho_{9,\rm c}}\right)^{4.3}
\left(\frac{7}{T_{8,\rm c}}\right)^{(\nu -3)} (\frac{0.5}{X_{12}})^2 \ 
{\rm s},
\end{split}
\end{equation}
where it is assumed that $L = dH/dt$ (see Table \ref{tab1}).

Since the central temperature, or the total heat content of the star
need only increase $\sim$3\% for the power to double, the white dwarf
evolution is governed by the rise of the power, $L$. Thus, the nuclear
timescale $\tau_{\rm L}$ and the time to explosion are the same and
both are considerably shorter, by about a factor of 5, than the
thermal time scale $\tau_{\rm H} \, = \ \left(\frac{\mathrm{d}\ln
H}{\mathrm{d}t}\right)^{\!\!-1}$.
The star takes a few centuries to go from ignition at $3.0 \times
10^8$ K to $7 \times 10^8$ K, but explodes a few minutes later. 
This is a considerably longer time scale than one would get by
evaluating a {\sl local} nuclear timescale $\tau_{\rm nuc} = (\cp T)/
(\nu \epsilon_{\rm nuc})$, just from the central energy
generation rate and heat capacity, because the cooler middle regions
of the white dwarf act as an appreciable heat reservoir for the energy
generation. Until the central temperature exceeds $7 \times 10^8$ K,
convection keeps the two efficiently coupled (see discussion in
\S4).

\clearpage

\begin{table*}
\centering
\caption{Nuclear and explosion timescales (see \S2)}
\begin{tabular}{lllllll}
\hline\noalign{\smallskip}
$T_{8,\rm c}$ & $\rho_{9,\rm c}$   & 
M$_{\rm conv}$  & $L$ & $\frac{dH}{dt}$  & 
$\tau_{\rm L}$  & 
$\texpl$ \\
&& ($\msol$) & (erg/s) & (erg/s) & (yr) & (yr) \\
\hline
3.0 & 3.60 & 0.2 & 2.7$\times10^{36}$ & 2.3$\times10^{36}$ & 210  & 300 \\
4.0 & 3.50 & 0.6 & 2.1$\times10^{39}$ & 1.9$\times10^{39}$ & 0.75  & 1.5 \\
5.0 & 3.35 & 0.85 & 3.3$\times10^{41}$ & 2.9$\times10^{41}$& 0.010  & 0.017 \\
6.0 & 3.16 & 1.05 & 1.9$\times10^{43}$& 1.6$\times10^{43}$& 3.5$\times10^{-4}$  & 4.6$\times10^{-4}$\\ 
\noalign{\smallskip}
\hline\noalign{\smallskip}
\end{tabular}\\
\flushleft
{NOTE: $T_{8,\rm c}$ is the central temperature in units of $10^8\,$K;
$\rho_{9,\rm c}$ is the central density in units of
10$^9\,$g$\,$cm$^{-3}$; M$_{\rm conv}$ the convective core;
$L$ the nuclear power defined in Eq.~(5); $\frac{dH}{dt}$ the
time derivative of the total heat content $H$ defined in Eq. (4);
$\tau_{\rm L}$ and $\texpl$ are respectively the nuclear and the
explosion timescales in years.}
\label{tab1}
\end{table*}

\clearpage
 
\subsection{Stellar Models}

Computer models of accreting white dwarfs approaching criticality were
constructed using the KEPLER stellar evolution code \citep{wzw78}.  A
composition of 30\,\% $^{12}$C and 70\,\% $^{16}$O, by mass fraction
was assumed, though the outcome will not depend appreciably on this
assumption. The initial model consisted of a ``warm'' white dwarf with
central temperature 10$^8$\,K and a mass of $2.6\times10^{33}$\,g
($\sim1.307\,\msol$).  Test calculations that started from a cooler
white dwarf resulted in a higher mass at ignition, but the results of
the pulsation analysis were not affected by this modification.  The
white dwarf then accreted matter (also carbon and oxygen in the same
proportion) at $10^{-7}$ $\msol\,$yr$^{-1}$.  The accretion was
stopped when the runaway had started, but about 80 centuries before
the final incineration, to allow the addition of the well-resolved
surface layers.  At this point the star has reached a central density
of $\rho_{9,\rm c}=3.64$ and a central temperature of $T_{8,\rm
c}=2.43$.  In the remaining time the star would have accreted less
than $10^{-3}\,\msol$, which does not affect the structure of the star
or the runaway (note that the runaway occurs many times that mass
before reaching the Chandrasekhar mass).

The Lagrangian grid uses zone masses ranging from 10$^{29}$\,g in the
center over some $10^{31}$\,g in the middle of the star to a smooth
gradient in zone size down to 10$^{15}$\,g at the surface.  Convection
is treated using standard mixing length theory with a mixing length
$L_{\rm mix} = H_{\rm P}$.

\begin{figure}
\psfig{file=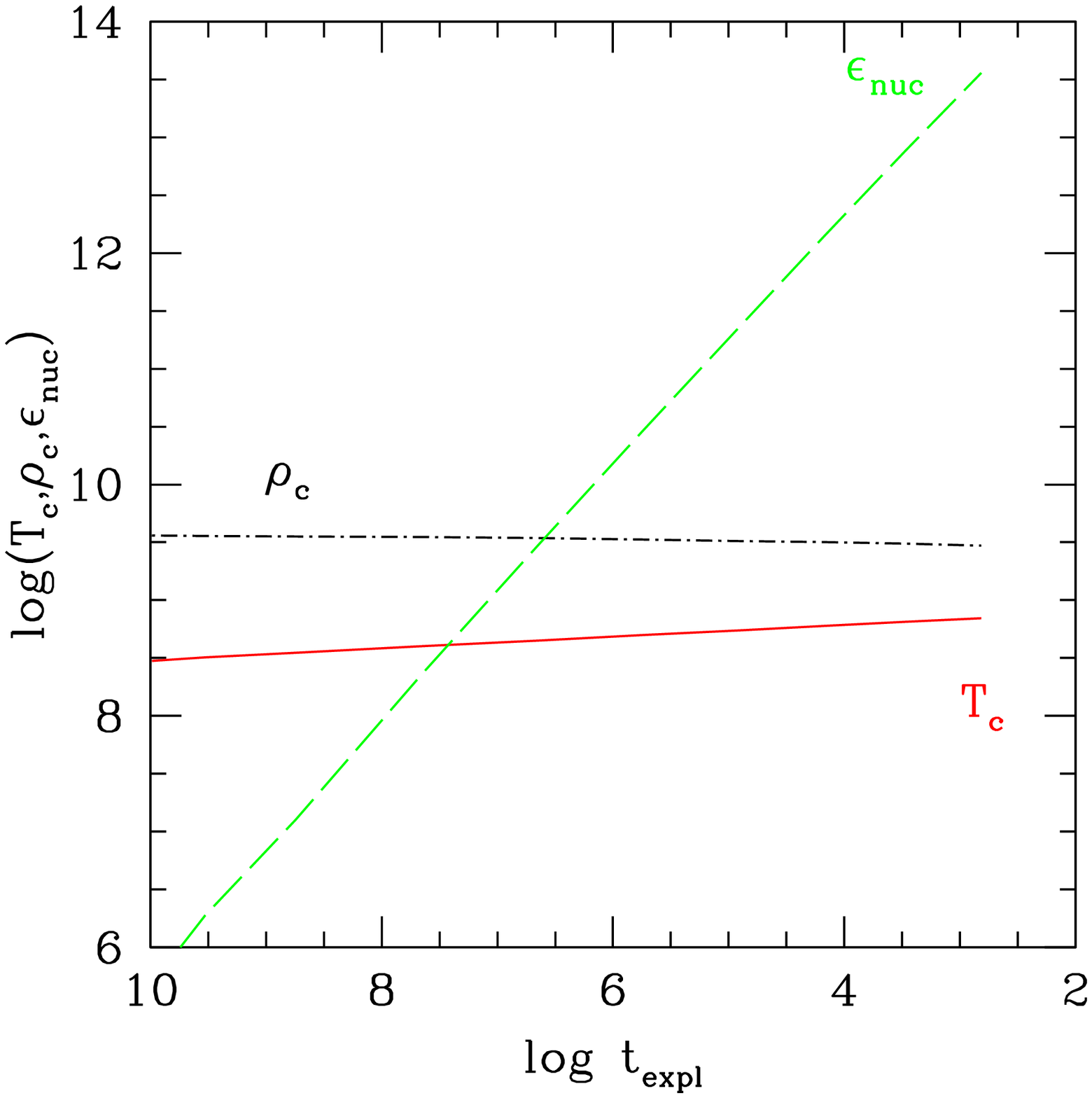,height=160mm,width=160mm} 
\caption{Evolution of the central temperature, $T_{\rm c}$, (in K; solid
line), up to 7 10$^8$K, central density $\rho_{\rm c}$ (in g ${\rm cm^{-3}}$;
dash-dotted line) and central nuclear energy $\enuc$ (in erg/g/s; dashed
line) of the white dwarf as a function of the remaining time before
explosion $\texpl$ (in s).}
\label{fig1}
\end{figure}

During the last $\sim$ 300\,yr prior to explosion the nuclear energy
generation rate, $\epsilon_{\rm nuc}$, rises rapidly in the center of
the white dwarf.  This is illustrated in Fig.~\ref{fig1} that displays
the evolution of central temperature, density and nuclear energy as a
function of the time remaining before explosion ($\texpl=0$
corresponds to the time of flame ignition/start of explosion).  While
only a small fraction of the energy released goes into expanding the
star, the expansion is significant because $\Gamma \approx 4/3$.


\section{Pulsation analysis}

The system of equations which are linearized are the basic
hydrodynamic equations written under the assumption of spherical
symmetry (cf. Unno et al. 1989, p.89). Details of the method and
pulsation code can be found in Baraffe et al.\ (2001). The
unperturbed, equilibrium state is described by hydrostatic
equilibrium. The nuclear energy term appearing in the energy
conservation equation includes neutrino energy loss.  
An important uncertainty in the present analysis is due to
the assumption that convection is frozen in, neglecting the perturbation 
of the convective flux. This simplification is based on the argument that
the convective timescales in the interior of the white dwarf remain
significantly larger than the period of pulsation of the excited modes
(see below).
A sequence of
evolutionary models is analyzed, starting from a central temperature
$T_{8,\rm c} = 3$, $\sim$ 300 yr before explosion, to $T_{8,\rm c} =
7$, corresponding to $\sim$ 10 minutes before explosion.  All models
analyzed show that the fundamental and higher order modes are
unstable.  As in Baraffe et al.\ (2001), the assumed time dependence
for the eigenfunctions is of the form $\exp({\rm i} \sigma t) \,\times
\,\exp(Kt)$, where $\sigma = 2\pi/\Pi$ is the eigen-frequency, $\Pi$
the corresponding period and $K$ the stability coefficient.  The
$e$-folding time characterizing the growth timescale of the pulsation
amplitude is defined by $\taud = 1/K$.  In all models analyzed, this
fundamental growth timescale is, by at least one order of magnitude,
shorter than $\taud$ of the higher order modes. Our analysis below can
thus be restricted to the fundamental mode, since it is the mode which
grows the fastest.

\begin{figure}
\psfig{file=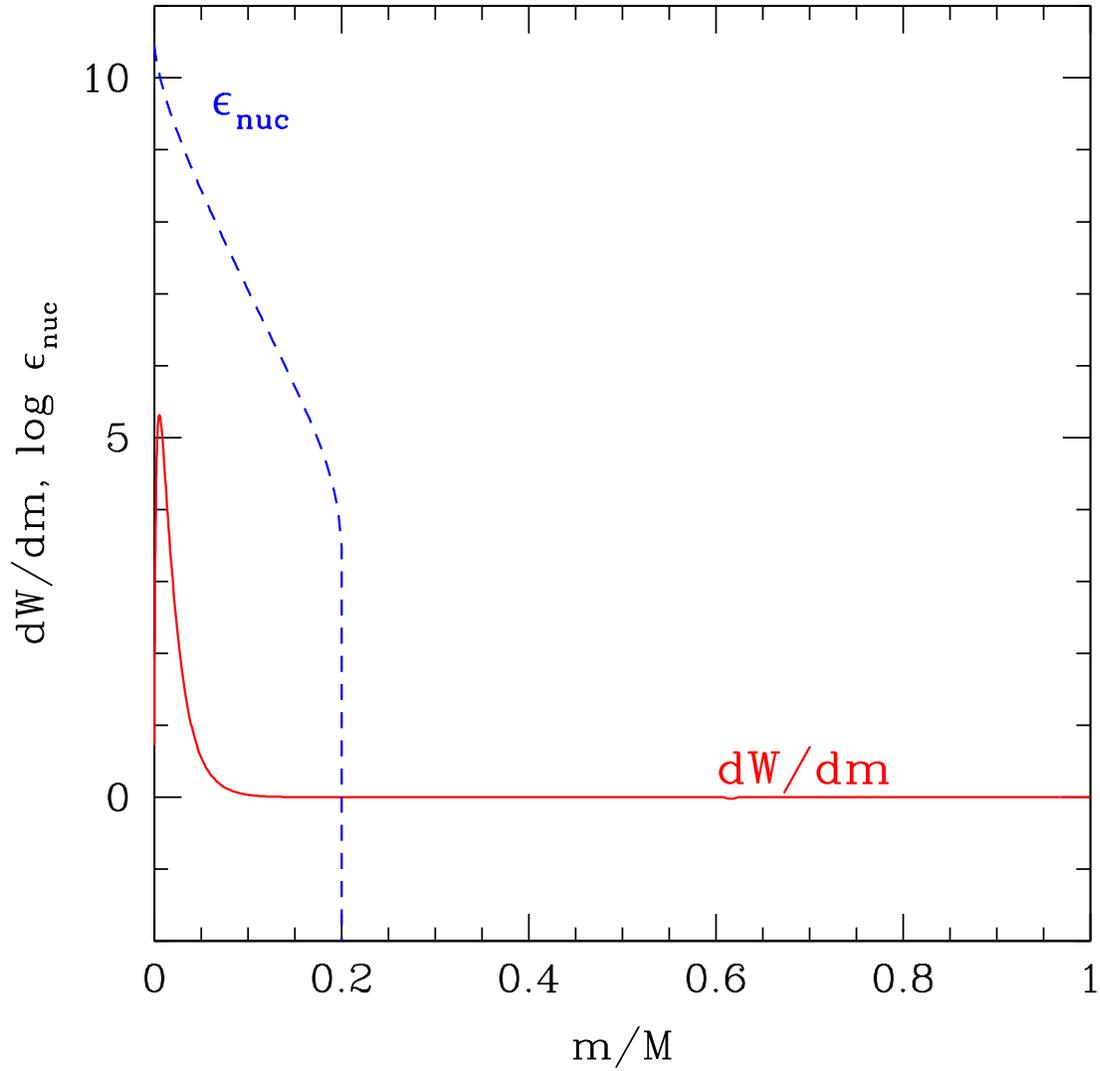,height=160mm,width=160mm} 
\caption{Differential work d$W$/d$M$, in arbitrary units, as a
function of mass (in units of the total mass) in the interior
structure of a model 6 days before explosion ($T_{8,\rm c} =
5$). The dashed line corresponds to the nuclear energy
generation $\log \, \enuc$ (erg/g/s).  }
\label{fig2}
\end{figure}

Table~2 summarizes the stability analysis results. The fundamental
mode pulsation period $\Pi_0$ is $\sim$ 2 s, close to the dynamical
timescale.  
 Note that the shortest convective timescale is $\sim \, 10^4$ s
in the first model analysed ($T_{8,\rm c}$ = 3) and $\sim \, 100$ s
in the last models.
For a model with $T_{8,\rm c} = 5$ ($\sim$ 6 days
before explosion), Fig. \ref{fig2} displays the work d$W$/d$M$,
which is proportional to the energy transferred to the pulsation,
with $W$ the so-called work integral (Cox 1980; Unno
et al.~1989).  Positive work indicates driving regions whereas
negative work corresponds to damping zones. The positive value of
d$W$/d$M$ in the center is due to the perturbation of the nuclear
energy generation rate
and characterizes the $\epsilon$-mechanism. As expected for a
gas with $\Gamma_1 \sim 4/3$, the displacement $\delta r/r$ is
essentially constant through the whole structure. Damping processes
are small in the rest of the white dwarf, and the total work integral
$W$ is essentially determined by the driving zone in the center.

\clearpage

\begin{table*}
\centering
\caption{Linear stability analysis results for an evolutionary
sequence of SN Ia progenitors prior to explosion.}
\begin{tabular}{lllllll}
\hline\noalign{\smallskip}
$T_{8,\rm c}$ & $\rho_{9,\rm c}$ & $R$ (Mm) & $\texpl$ (s) &
$\taunuc$ (s) & $\taud$ (s) & $\Pi_0$ (s)\\
\noalign{\smallskip}
\hline\noalign{\smallskip}
3  & 3.60  & 1.75 & 9.68$\times10^9$ & 2.99$\times10^8$ & 2.95$\times10^{11}$ & 1.95\\
4  & 3.50  & 1.77 & 4.48$\times10^7$ & 8.31$\times10^5$ & 6.19$\times10^{8}$ & 1.95\\
5  & 3.35  & 1.79 & 5.46$\times10^5$ & 8.63$\times10^3$ & 4.92$\times10^{6}$ & 1.96\\
6  & 3.16  & 1.82 & 1.47$\times10^4$ & 2.26$\times10^2$ &  1.03$\times10^{5}$ & 1.97\\
7  & 2.97  & 1.85 & 6.57$\times10^2$ & 11.3 & 4.22$\times10^{3}$ & 1.98\\
\hline
\end{tabular}\\
\flushleft
{NOTE: Models are characterized by the central temperature $T_{8,\rm
c}$, central density $\rho_{9,\rm c}$ and white dwarf radius $R$.
$t_{\rm expl}$ is the remaining time before explosion and $\taunuc =
\cp T/(\nu \enuc)$ is the local nuclear timescale defined in the
center.  Growth timescale 
(see text) and period of the fundamental mode are denoted respectively by
$\taud$ and $\Pi_0$.}
\label{tab2}
\end{table*}

\clearpage


\section{Discussion}

Linear stability analysis alone does not provide information about the
maximum pulsation amplitude which can be finally reached. For this
purpose, full hydrodynamical calculations are needed.  However,
comparison between the growth timescale $\taud$ of an excited mode and the
evolutionary timescale $\texpl$ can indicate whether the perturbation
may have time to grow in order to reach large amplitudes.  Before
doing such comparisons, it is useful to consider the uncertainties
affecting the evolutionary timescale.


As previously discussed in \S2.1, the relevant timescale
characterizing the evolution to explosion is considerably longer than
the local nuclear timescale because of the heat reservoir provided by
convective coupling to the rest of the star. 
For an $\epsilon$-mechanism, the growth timescale is expected
 to be closely
related to the local nuclear timescale, since $\taunuc$ characterizes
the increase of local nuclear energy generation rate in the central region
where the $\epsilon$-mechanism takes place (see Fig. \ref{fig2}).
The discussion below  will thus be based on $\taunuc$ and its 
comparison with the time
to explosion. Qualitatively, convection can lengthen the evolutionary
timescale compared to the local nuclear timescale, and thus $\taunuc
\le \texpl \approx \tau_{\rm L}$.

\begin{figure}
\psfig{file=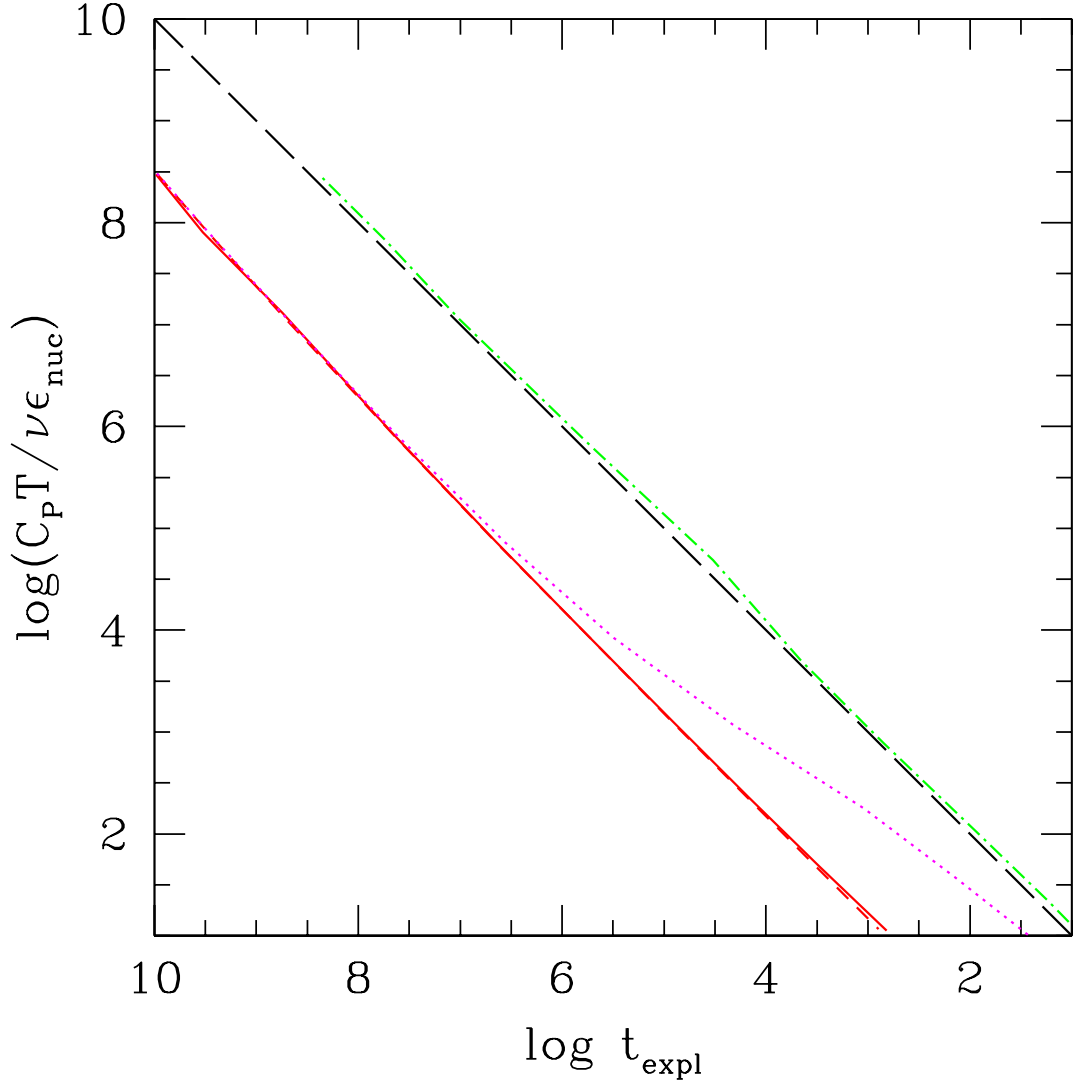,height=160mm,width=160mm} 
\caption{Local nuclear timescale $\taunuc$ (see text) as a function of
the remaining time before explosion $\texpl$ (in s) for different
cases of convection efficiency: $L_{\rm mix} = H_{\rm P}$ (solid
line), $L_{\rm mix} = 100 H_{\rm P}$ (dashed line, almost indistinguishable from the solid line), $L_{\rm mix} =
H_{\rm P}/100$ (dotted line) and $L_{\rm mix} =0$ (dash-dotted
line). The long-dashed line indicates the case where $\taunuc$ =
$\texpl$.  }
\label{fig3}
\end{figure}

For the sake of a quantitative comparison, we have performed test
evolutionary sequences with different convection efficiencies, in
terms of the mixing length parameter $L_{\rm mix}$. All test sequences
start from the same model with central temperature $T_{8,\rm c} = 3$.
The nuclear timescale $\tau_{\rm nuc}$ as a function of the remaining
time before explosion is displayed in Fig.~\ref{fig3} for four
different cases of convection efficiency: the standard case with
$L_{\rm mix} = H_{\rm P}$ (solid line), $L_{\rm mix} = 100 H_{\rm P}$
(dashed line), $L_{\rm mix} = H_{\rm P}/100$ (dotted line) and a case
with no convection (dash-dotted line).  The long-dashed line
corresponds to the case $\texpl = \taunuc$.  As expected, when
convection is inhibited, the runaway timescale is essentially
determined by the local nuclear timescale. On the opposite, the more
efficient the convection, the longer the evolutionary timescale
compared to $\taunuc$.  Fig. \ref{fig3} indicates that even in the
unrealistic case with $L_{\rm mix} = 100 H_{\rm P}$, the evolutionary
timescale is at most $\sim$ 70 times greater than
$\taunuc$. Therefore, although the treatment of convection is crucial
and still very uncertain during these last phases of evolution, one
can reasonably expect that such uncertainty will not lengthen the
evolutionary timescale by more than $100 \times \taunuc$.



\begin{figure}
\psfig{file=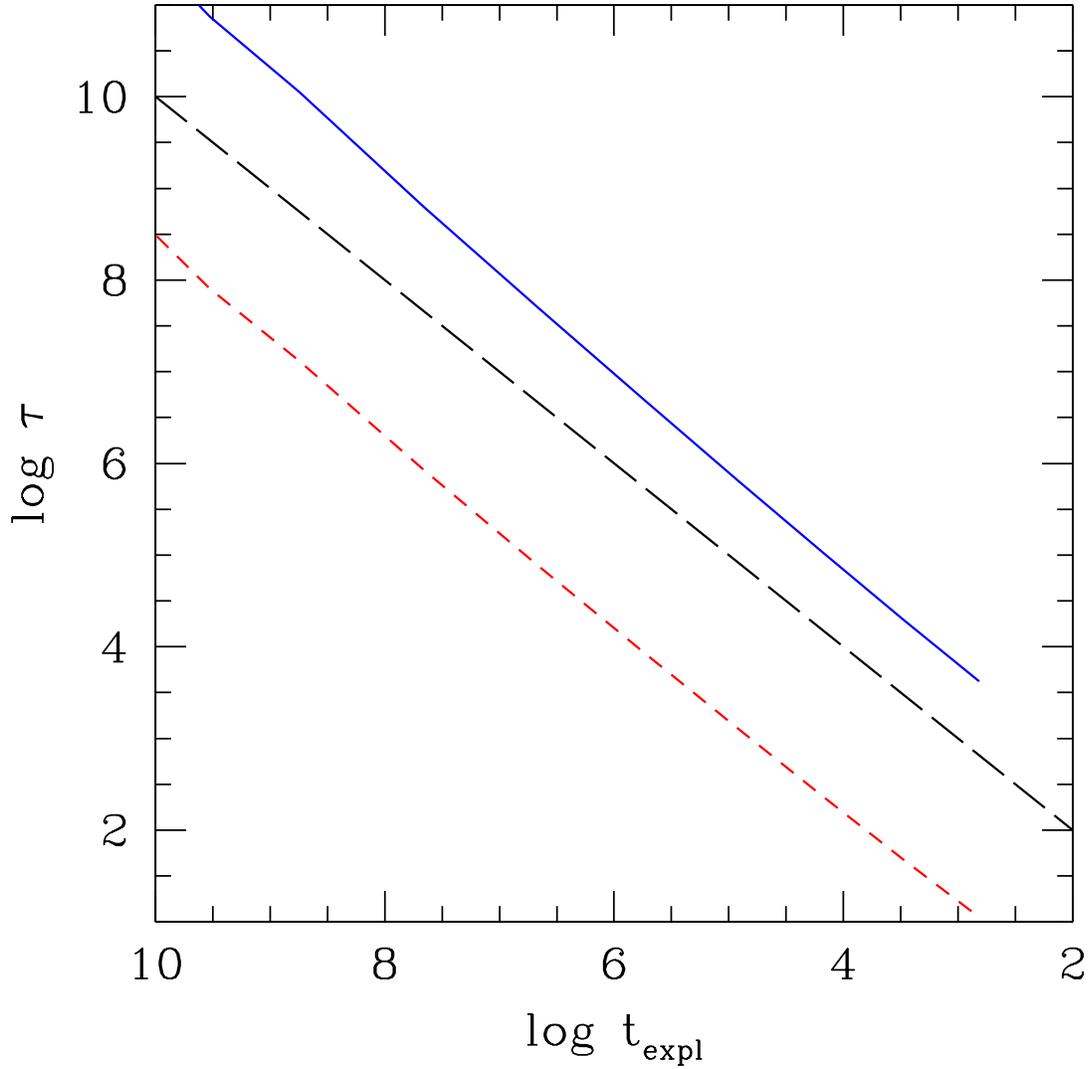,height=160mm,width=160mm} 
\caption{Characteristic timescales (in s) as a function of the
remaining time before explosion $\texpl$ (in s): growth timescale for
the vibrational instability $\taud$ (solid line) and local nuclear timescale
$\taunuc$ (short-dashed line). 
The long-dashed line indicates the ideal
case where the characteristic timescales are equal to $\texpl$.  }
\label{fig4}
\end{figure}

Fig.~\ref{fig4} displays  the growth timescale
$\taud$ of the fundamental mode and the local nuclear timescale $\taunuc$ 
as a function of the evolutionary timescale
$\texpl$, for the standard case of convection efficiency
(see also Table \ref{tab2}).
As expected and illustrated in Fig.~\ref{fig4}, 
$\taud$ follows $\taunuc$, with $\taud$/$\taunuc \, \sim 5\, 10^2 - 10^3$,
the same relation of proportionality being found independently of 
convection efficiency.
Since the tests on convection efficiency suggest that $\texpl$ $<$ 
100 $\taunuc$, the quantity $\taud$ thus remains
systematically larger than the evolutionary timescale $\texpl$, 
as illustrated in Fig. \ref{fig4}.
For standard convection, $\taud$ is larger than $\texpl$
by a factor
$\sim$ 30 at the beginning of the sequence ($T_{8,\rm c} = 3$) down to
a minimum factor of $\sim$ 6.5 for $T_{8,\rm c} = 7$.  Similar
behavior is found for the test sequences with varying convection
efficiency, with the smallest ratio $\taud/\texpl \sim 5$ found for
the case $L_{\rm mix} = 100 H_{\rm P}$.  Such results suggest that the
present vibrational instability does not have time to grow to reach
significant amplitudes.

\section{Conclusion}

Based on a linear stability analysis of Chandrasekhar-mass white dwarf
models prior to explosion, we have found that the central conditions
are favorable to the excitation of radial eigenmodes through the
$\epsilon$-mechanism.  Such pulsational instability could in principle
result in phases of expansion and contraction with growing amplitude
and thus affect the structure of the supernova progenitor.  We also
found, however, that the growth timescale for the pulsation amplitude
remains systematically longer than the evolutionary timescale.
 Such instability does not have time to grow significantly before the
explosion of the white dwarf.

\acknowledgements

AH performed this work under the auspices
of the U.S.\ Department of Energy at the Los Alamos National
Laboratory operated by the University of California under contract
No.\ W-7405-ENG-36, in part through support by the Department of
Energy under grant B341495 to the Center for Astrophysical
Thermonuclear Flashes at the University of Chicago, and by support
through a Fermi Fellowship at the University of Chicago.  The research
of SW in this area is supported by the NSF (AST 02-06111), NASA
(NAG5-12036), and the DOE SciDAC Program under grant DE-FC02-01ER41176.

{}

\clearpage

\vfill\eject

\vfill\eject

\vfill\eject

\end{document}